\documentclass[twocolumn,amsmath,amssymb, superscriptaddress]{revtex4}

\usepackage{graphicx}
\usepackage{dcolumn}
\usepackage{bm}
\usepackage{epstopdf}
\usepackage{color}

\begin{document}


\title{Single-bubble and multi-bubble cavitation in water triggered by \\laser-driven focusing shock waves}

\author{D. Veysset}
\affiliation{Department of Chemistry, Massachusetts Institute of Technology, Cambridge, MA 02139, USA}
\affiliation{Institute for Soldier Nanotechnologies, Massachusetts Institute of Technology, Cambridge, MA 02139, USA}

\author{U. Guti\'errez-Hern\'andez}
\affiliation{Instituto de Ciencias Nucleares, Universidad Nacional Aut\'{o}noma de M\'{e}xico, Apartado Postal 70-543, 04510 Cd. Mx., M\'{e}xico}

\author{L. Dresselhaus-Cooper}
\affiliation{Department of Chemistry, Massachusetts Institute of Technology, Cambridge, MA 02139, USA}
\affiliation{Institute for Soldier Nanotechnologies, Massachusetts Institute of Technology, Cambridge, MA 02139, USA}

\author{F. De Colle}
\affiliation{Instituto de Ciencias Nucleares, Universidad Nacional Aut\'{o}noma de M\'{e}xico, Apartado Postal 70-543, 04510 Cd. Mx., M\'{e}xico}

\author{S. Kooi}
\affiliation{Institute for Soldier Nanotechnologies, Massachusetts Institute of Technology, Cambridge, MA 02139, USA}

\author{K. A. Nelson}
\affiliation{Department of Chemistry, Massachusetts Institute of Technology, Cambridge, MA 02139, USA}
\affiliation{Institute for Soldier Nanotechnologies, Massachusetts Institute of Technology, Cambridge, MA 02139, USA}

\author{P. A. Quinto-Su}
\email{pedro.quinto@nucleares.unam.mx}
\affiliation{Instituto de Ciencias Nucleares, Universidad Nacional Aut\'{o}noma de M\'{e}xico, Apartado Postal 70-543, 04510 Cd. Mx., M\'{e}xico}

\author{T. Pezeril}
\email{thomas.pezeril@univ-lemans.fr}
\affiliation{Institut Mol\'ecules et Mat\'eriaux du Mans, UMR CNRS 6283, Universit\'e du Maine, 72085 Le Mans, France}

\begin{abstract}

In this study a single laser pulse spatially shaped into a ring is focused into a thin water layer, creating an annular cavitation bubble and cylindrical shock waves: an outer shock that diverges away from the excitation laser ring and an inner shock that focuses towards the center. A few nanoseconds after the converging shock reaches the focus and diverges away from the center, a single bubble nucleates at the center. The inner diverging shock then reaches the surface of the annular laser-induced bubble and reflects at the boundary, initiating nucleation of a tertiary bubble cloud. In the present experiments, we have performed time-resolved imaging of shock propagation and bubble wall motion. Our experimental observations of single-bubble cavitation and collapse and appearance of ring-shaped bubble clouds are consistent with our numerical simulations that solve a one dimensional Euler equation in cylindrical coordinates. The numerical results agree qualitatively with the experimental observations of the appearance and growth of {\color{black}large} bubble clouds at the smallest laser excitation rings. Our technique of shock-driven bubble cavitation opens novel perspectives for the investigation of shock-induced single-bubble or multi-bubble {\color{black}cavitation} phenomena in thin liquids.

\end{abstract}

\maketitle


\section{Introduction}

Liquids can withstand tensions due to intermolecular attractive potentials up to a tensile limit above which liquids rupture and bubbles nucleate. This limit varies depending on many factors, including the nature of the liquid, the purity of the liquid, as  impurities drastically lower the limit through heterogeneous nucleation, the characteristics of the container \cite{herbert2006}, and the rate at which the tensile force is applied \cite{maxwell2013, arvengas2011, ando2012}. Water, in particular, has a wide range of measured tensile limits (from few to hundred negative MPa) that depend strongly on the rate at which the tensile force is applied \cite{maxwell2013, arvengas2011, ando2012}. Bubble generation, or cavitation, upon liquid rupture has implications in a variety of areas in technology and fundamental science. For instance, cavitation has been proposed as a damage mechanism for traumatic blast injury but the phenomenon is still poorly understood due to the difficulty for real-time observations \cite{Goeller}. In addition, the extreme conditions of pressure and temperature reached during bubble collapse are of great interest for chemists that investigate chemical reactions under the influence of sound \cite{Suslick}. There is therefore a clear need for reproducing the conditions for reliable bubble cavitation in the laboratory to allow systematic observations and studies of cavitation phenomena. {\color{black} Cavitation bubbles can be} generated upon reflection of shock waves at liquid-gas {\color{black} or liquid-solid boundaries \cite{ando2012, Seddon2012, Ando2012}} or upon interaction of shock waves \cite{quinto2013, ohl2013}. In the present work we pursue the shock-focusing configuration introduced by Pezeril et al. where a picosecond laser pulse shaped into a ring \cite{pezeril2011, veysset2015, veysset2016, veysset2017} is focused into a thin absorbing liquid sample to create high amplitude converging shock waves and cavitation bubbles. {\color{black} This configuration enables the generation of localized high pressure away from the laser focus, contrarily to classical laser cavitation experiments \cite{ohl1999, Bell, Vogel}, and the real-time observation of propagating shock waves as well as dynamics of cavitation bubbles. We experimentally evidence several stages and pathways of laser shock-induced cavitation phenomena. First, we observe the dynamics of the laser-induced annular cavitation bubble, coupled to the onset of the laser shock excitation. Second, a few nanoseconds after the shock converges at the center of the ring, a bubble nucleates at the focus as the shock rebounds and diverges away from the center. Third, in addition to the central bubble dynamics, we also observe the inner shock reflection at the annular laser-induced bubble and the subsequent nucleation of a tertiary bubble cloud.}

In the present work, we expand the initial study on laser ring excitation \cite{pezeril2011, veysset2015, veysset2016} by exploring longer time delays in order to observe the dynamics of nucleated bubbles. We also model the shock wave propagation and focusing to quantify the effect of the laser ring radius on the negative pressure reached at the center. The paper is organized in the following way. First, we describe the experimental setup and show the results of bubble nucleation from single-shot experiments. Second, we study the effects of varying the laser ring radius on bubble nucleation using stroboscopic imaging. Finally, we discuss our numerical simulations on the shock dynamics obtained from a one-dimensional axisymmetric Euler solver.



   \begin{figure}[t!]
   \includegraphics[width=8.5cm]{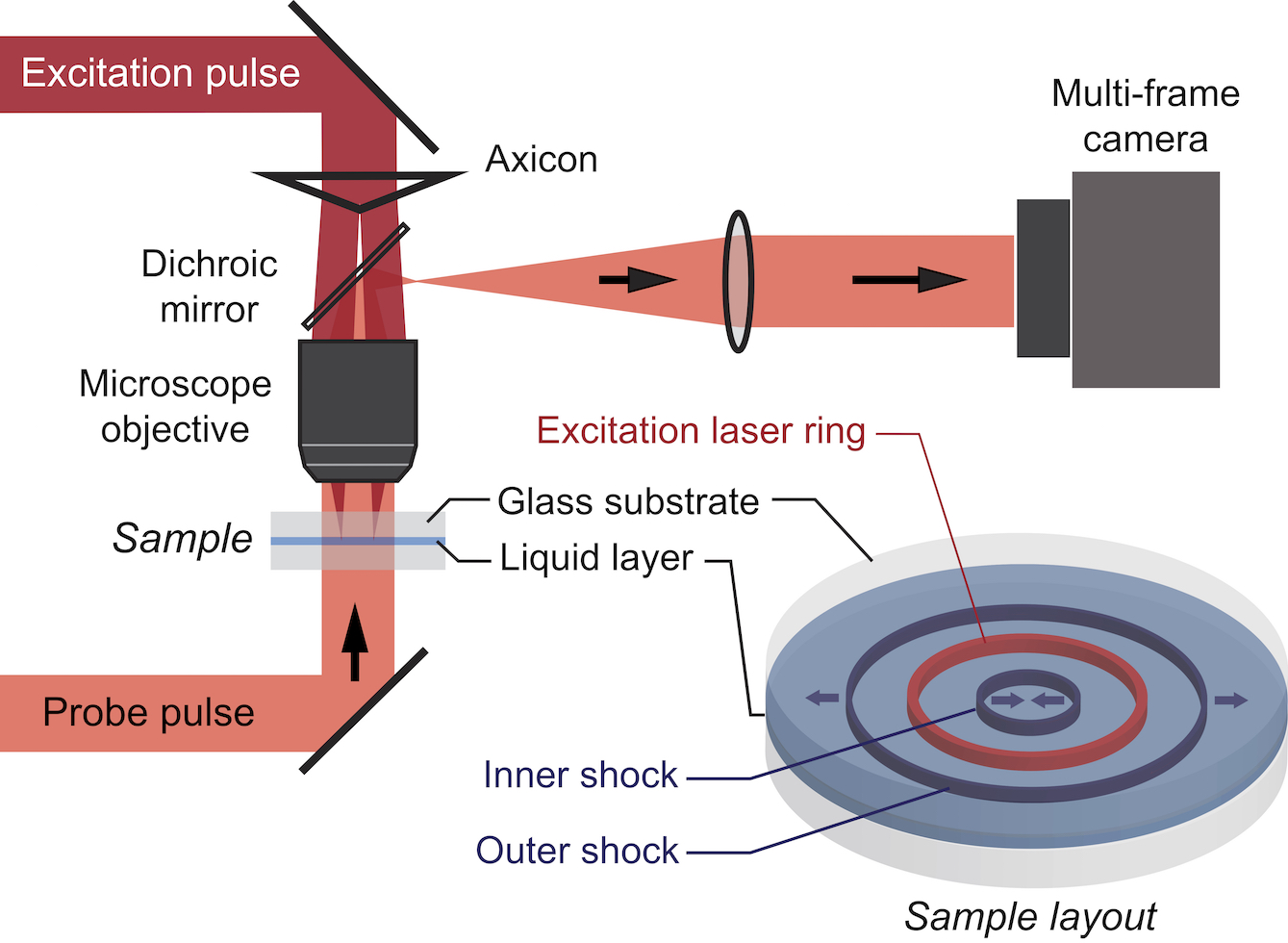}
   \caption[Fig1]{Experimental setup. An axicon combined with a lens is used to focus a laser excitation pulse as a ring at the sample location. The sample is illuminated by a probe pulse and imaged with high magnification on a multi-frame camera. After laser absorption by the liquid sample, two in-plane counter-propagating shock waves are launched and remain mostly confined within the liquid layer.}
   \label{Fig1}
   \end{figure} 

\section{Single shot multi-frame imaging, experimental setup and results}

The experimental setup is depicted in Fig.~\ref{Fig1}(a). A 150-ps duration, 800-nm wavelength, laser pulse delivered by an amplified Ti:sapphire system is focused into a 10 $\mu$m-thick liquid layer as described in \cite{pezeril2011, veysset2015, veysset2016}. The thin liquid layer consists of a suspension of carbon nanoparticles in water (India ink diluted to yield 2$\%$ weight carbon concentration). {\color{black}The carbon concentration was chosen to find a good compromise between sufficient pump absorption for efficient shock generation and sufficient probe transmission for bright imaging.} The layer is confined between two glass windows separated by a polymer spacer. The laser excitation pulse is shaped into a ring of 95$~\mu$m in radius in the plane of the liquid layer using a $0.5^{\circ}$ axicon and a 3~cm focal length achromatic doublet as sketched in Fig.~\ref{Fig1}(a). After each shot of the laser excitation pulse, the sample is moved using a motorized stage to an undisturbed area in order to avoid remnant bubbles. 

The time-resolved images are obtained through high-speed imaging. The high-frame-rate camera (SIMX 16, Specialized Imaging) that is used in the experiment can acquire 16 frames on a single shot, with tunable exposure time and tunable time interval between frames. As an illuminating probe, we use a 640~nm wavelength laser (Cavilux, Cavitar Ltd) of 30$~\mu$s pulse duration, which is longer than the total time required to acquire the 16 frames on the high-frame-rate camera.

{\color{black}Flash heating of the carbon nanoparticles upon laser irradiation causes the water to vaporize thus to quickly expand, launching two counter-propagating shock waves propagating laterally within the liquid layer. In the present experiments, the stress generation through the vaporization process dominates over the thermoelastic process \cite{sigrist} that can be neglected.} The inner-propagating wave converges towards the center while the outer-propagating wave diverges, as sketched in Fig.~\ref{Fig1}(b). The converging shock accelerates upon convergence and increases in amplitude as it focuses towards the center of the ring. The diverging shock decreases in strength because of the combined effects of cylindrical divergence and attenuation. The rather efficient shock confinement within the liquid layer is ensured by the acoustic impedance mismatch between the liquid and the solid glass substrates \cite{Leora}.

   \begin{figure*}[t!]
   \includegraphics[width=\textwidth]{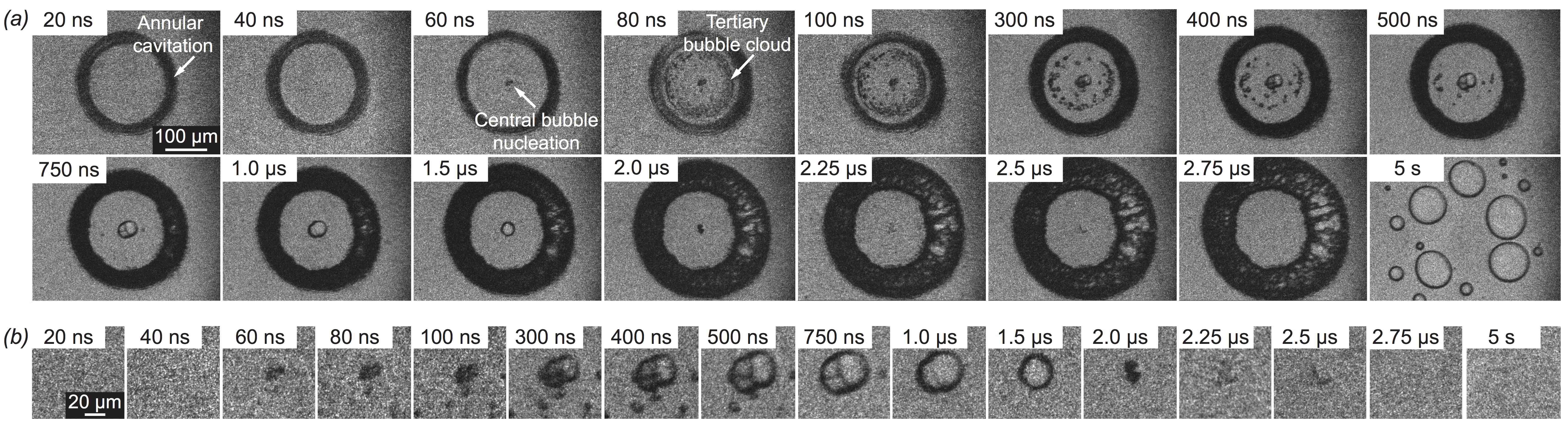}
   \caption[Fig2]{(a) Single-shot frames recorded for an excitation pulse of 0.5 mJ and a laser ring of 95$~\mu$m radius. The bubble formation at the vicinity of the annular bubble and the collapse dynamics of the central bubble are clearly apparent on this sequence of time resolved images. (b) Zoomed in images taken from (a) highlighting the dynamics of the central bubble.}
   \label{Fig2}
   \end{figure*}
   
Figure 2 shows a representative sequence of frames illustrating the bubble dynamics, recorded using a laser excitation energy of 0.5~mJ and a ring radius of 95~$\mu$m. The exposure time of each frame is set to 5 ns for frames 1 to 6 and 10 ns for the following frames. The shocks are not visible in these images, but based on previous work \cite{pezeril2011} on shock trajectories in this configuration, we expect the shock to reach the focus within 50~ns. At 60~ns, about 10~ns after shock focus, we observe the onset of bubble cavitation and growth at the center, while the rebounding inner diverging shock diverges toward the annular laser-induced bubble. {\color{black}The central bubble nucleation at the shock focus is a consequence of the Gouy phase shift, a well-known occurrence that has been observed through imaging of converging electromagnetic or acoustic waves \cite{pezeril2011}}. The following five frames show the appearance and evolution of a nucleated tertiary bubble cloud due to the inner shock being reflected at the annular laser-induced bubble. Finally, the tertiary bubbles disappear within a few hundreds of nanoseconds whereas the central bubble collapses in a timescale of 1-2 microseconds. {\color{black}Eventually, as shown in Fig. 2(a), at longer times (milliseconds to seconds), the annular bubble separates into several cylindrical bubbles. The fast collapse of the tertiary bubbles of very small diameters, in the order of 10 microns or smaller, suggests that those bubbles remain in the bulk of the water and are spherical, see supplemental where we have added a plot showing the collapse of the tertiary bubbles that suggests the 3D nature of those bubbles. As expected from the reflection of the shock pulse at the annular bubble wall, the bubble cloud appears at a distance from the annular bubble that corresponds to about half the pulse length. This is a consequence of the partial overlap of the shock front with its inverted part at the bubble boundary during the reflection process. This effect vanishes the effective tensile pressure jump at the vicinity of the boundary. For this reason, as seen in Fig. 2(a), the bubbles appear at a distance of about 10 $\mu$m away from the annular bubble.}

   \begin{figure}[tb!]
   \includegraphics[width=8.5cm]{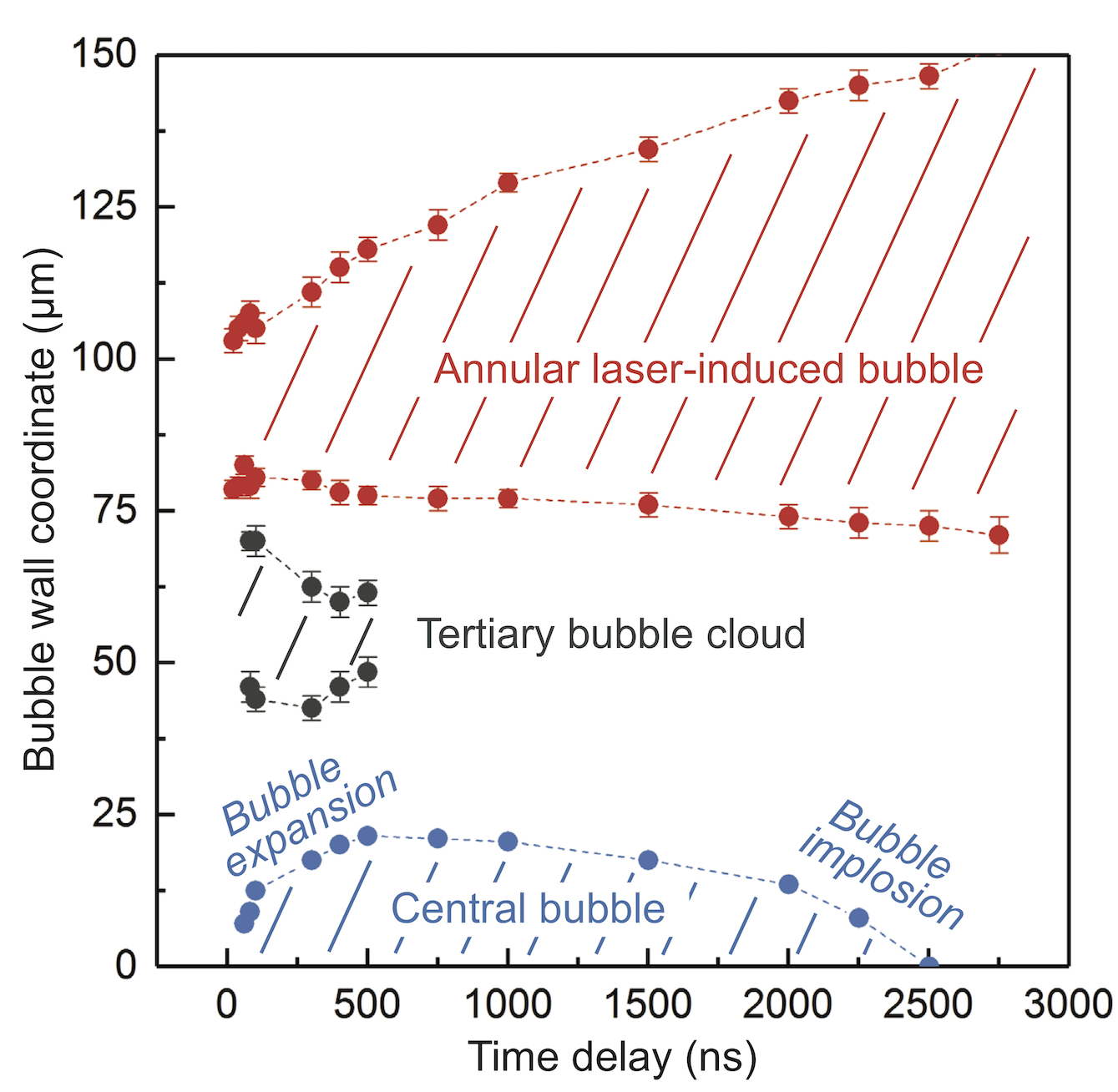}
   \caption[Fig3]{Bubble walls trajectories extracted from the frames displayed in Fig.~2. The two stages of the central bubble wall trajectory (expansion and collapse) have analogies with the classical Rayleigh-Plesset bubble dynamics well-established in single bubble sonoluminescence \cite{brenner}.}
   \label{Fig3}
   \end{figure}
   
To obtain the trajectories of the bubble walls, we extract the positions of the bubble boundaries from each of the frames presented in Fig.~\ref{Fig2}. The extracted trajectories are shown in Fig.~\ref{Fig3}. From the trajectories of the bubble walls, in particular the trajectory of the central bubble, we estimate the bubble wall average speed during the bubble expansion (first stage) and during the bubble collapse (second stage). The two stages of bubble motion have noticeable different average speeds, $v_{expansion}\sim$ 40 $\mu$m/$\mu$s and $v_{collapse}\sim$ 10 $\mu$m/$\mu$s. The asymmetry in the bubble wall motion during each of the expansion and collapse stages calls for a comparison with the classical single bubble wall motion in a liquid driven by an external acoustic field. In the latter situation, the non-linear 3D Rayleigh-Plesset equations models accurately the first stage of smooth expansion and the second stage of violent collapse of the single bubble motion, which leads to the observation of sonoluminescence \cite{brenner}. As a comparison, the average bubble wall speed in sonoluminescence is in the range of $v_{expansion}$ = 3 $\mu$m/$\mu$s and $v_{collapse}$ = 10 - 20 $\mu$m/$\mu$s. Our results suggest that in the present experimental situation the first stage of bubble expansion is more violent than the second stage of bubble collapse, which does not match the well-established Rayleigh-Plesset 3d model used in sonoluminescence but is consistent with the slower collapse of bubbles in thin liquids modeled with a 2D Rayleigh-Plesset equation \cite{2d}. Hence, these estimations indicate that the conditions for the observation of the sonoluminescence phenomena are probably not fulfilled here. Even though our experimental attempts to observe sonoluminescence in our specific cylindrical configuration of bubble implosion were unsuccessful so far, these results of shock-driven bubble creation and implosion {\color{black}arise interest in the context of the sonoluminescence phenomena.} 

{\color{black}As it can be seen in Fig. 2(b), it appears that the central bubble nucleation occurs at multiple sites, which is particularly evident after 300~ns. This is not surprising and this is caused by the conjugation of two factors. First, because of experimental imperfections, the laser focus is not an ideal ring with homogeneous laser intensity distribution and consequently the shock focuses in a complex geometrical shape leading to several tensile sites for nucleation. Second, impurities, such as carbon nanoparticle clusters or trapped bubbles, and the liquid-solid interface act as multiple sites and opportunities for heterogeneous nucleation. If no particular effort is taken during sample preparation to eliminate impurities or treat the container surfaces, heterogeneous nucleation occurs at negative pressures inferior to the homogenous nucleation threshold (tensile strength of the liquid). For instance, the theoretical tensile strength of pristine water is about 140 MPa at 25$^{\circ}$C \cite{Fisher} but heterogeneous nucleation thresholds have been measured at varying pressures ranging from a few MPa to a few tens of MPa.}

   \begin{figure*}[t!]
 \includegraphics[width=12cm]{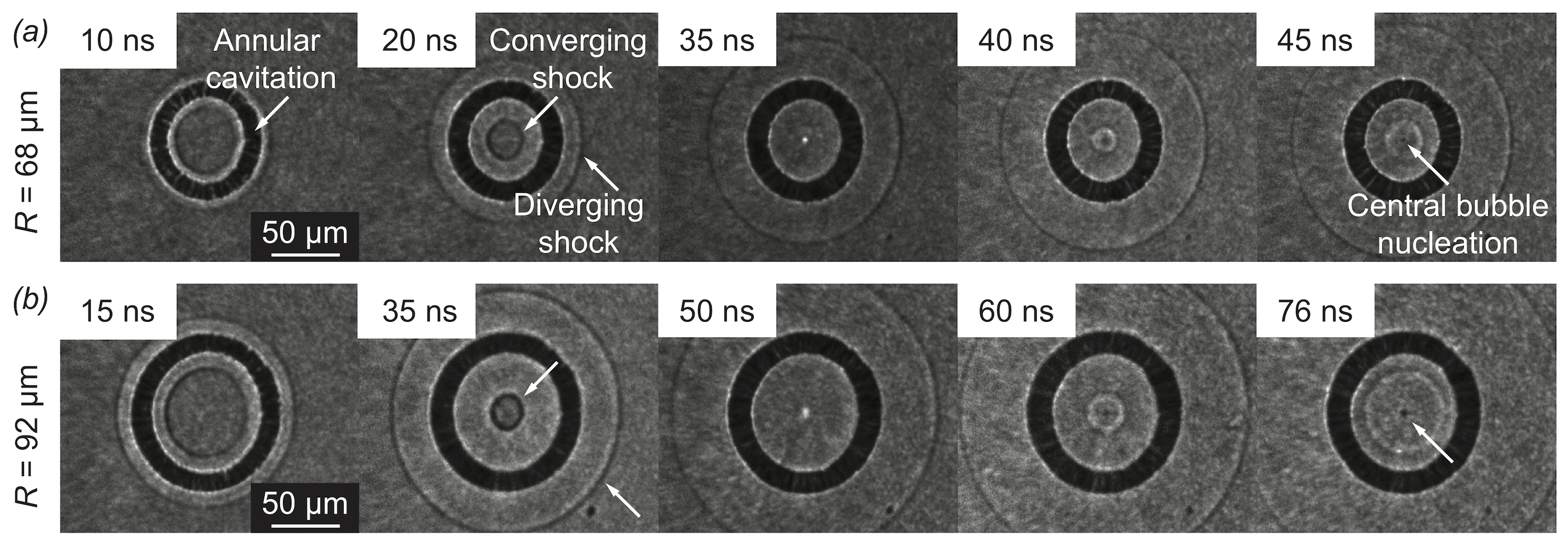}
    \caption[Fig4]{Selection of single-shot stroboscopic time-resolved images recorded for different laser ring radii $R$ = 68 ~$\mu$m (a), 92~$\mu$m (b). The central image taken at 35~ns and 50~ns in each sequence indicate the instant of inner shock focus. Each individual frame has a width of 410~$\mu$m. See supplemental for more radii.}
    \label{Fig4}
   \vspace{1cm}
   \includegraphics[width=12cm]{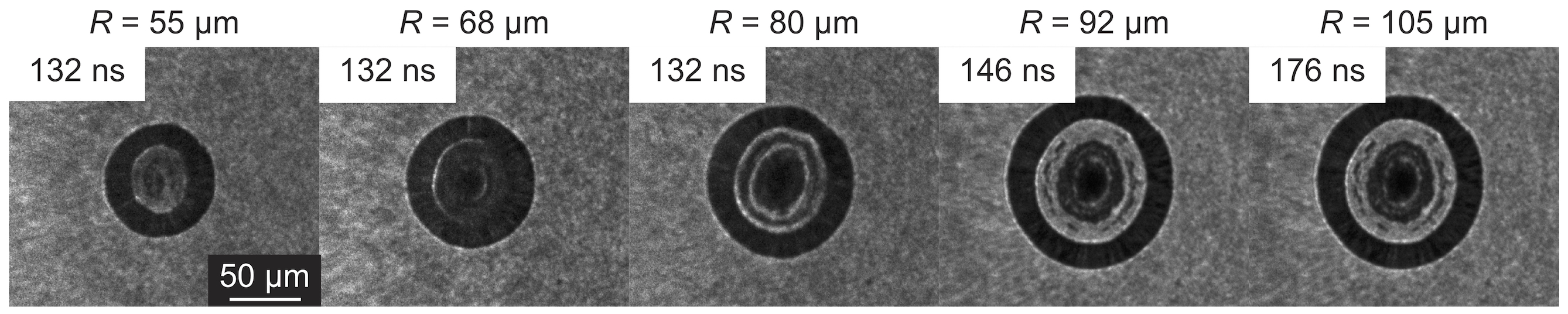}
   \caption[Fig5]{Selection of stroboscopic time-resolved images recorded such as in Fig. 4, at different radius, at times where the bubble cloud is the largest. The inner bubble cloud appears after reflection of the inner shock at the annular laser-induced bubble. Qualitatively, the bubble cloud appears to fill most of the inner part of the annular bubble at small laser ring radius. Each individual frame has a width of 410~$\mu$m.}
     \label{Fig5}
   \end{figure*} 

\section{Single shot stroboscopic imaging, experiments varying the laser ring radius}

Experiments varying the laser excitation ring radius $R$ were performed to observe the effect of radius on the tertiary bubble cloud for a constant laser excitation {\color{black} fluence of 25~J/cm$^{2}$}. Such single-shot experiments were performed on a stroboscopic manner {\color{black}using delayed flash illuminations} from two distinct electronically time delayed laser systems. In these stroboscopic experiments, the dynamics are imaged with strobe photography by changing the time delay between the excitation pulse and the imaging pulse. The experimental setup for the stroboscopic measurements is described in \cite{quinto2013} and it is similar to that shown in Fig.~\ref{Fig1}. For the excitation pulse, we used a Nd:YAG laser (New Wave, Solo PIV) with a duration of 6~ns and a wavelength of 532~nm. The excitation laser pulse is shaped with a computer-controlled spatial light modulator (SLM) to allow a straightforward modification of the laser ring radius at the focus of a 10~$\times$, 0.4~NA microscope objective. A second frequency-doubled Nd:YAG laser (New Wave, Solo PIV) is used for delayed flash illumination at a well defined time delay to capture the dynamics of the events. The beam is focused into a dye cell (Exciton Rhodamine 698 diluted in methanol) to excite emission centered at 698~nm. The emitted 698~nm imaging pulses are coupled to an optical fiber that directs the light into the condenser of the microscope to illuminate the sample. The single-shot events are imaged with an ICCD camera (Andor, IStar). The studied liquid is undiluted ink (T6643 Epson, magenta, $<80\%$ water by weight, 1.08 relative density) of 19~$\mu$m thickness confined in between two glass substrates. The studied laser ring radii $R$ are 55, 68, 80, 92, and 105~$\mu$m. {\color{black}In term of laser pulse energy, the diffracted energy at the largest radius of 105~$\mu$m is 222~$\mu$J. The energy for the other radii are adapted to maintain a constant laser fluence of 25~J/cm$^{2}$, assuming a laser ring width of 1.6~$\mu$m close to the diffraction-limited spot size. Similarly to the multi-frame imaging experiments, the Rayleigh length for the laser beam waist at focus is several times bigger than the sample thickness. The sample location is adjusted to the best focus with a tolerance of $\pm$ 10 microns.}

Figure~\ref{Fig4} shows the dynamics of the annular laser-induced bubbles and the shock-induced central bubbles for two distinct laser ring radius $R$ of 68 and 92~$\mu$m. The central images in Fig.~\ref{Fig4} shows the inner shock that focuses at the center of the laser ring. As expected, the time for shock focusing increases when the laser ring radius $R$ increases. At longer time delays, the diverging inner shock and a cavitation bubble growing at the center can be seen on the images displayed in Fig.~\ref{Fig4}. Once the inner diverging shock reaches the wall of the annular laser-induced bubble, it is reflected as a tensile shock wave and travels back toward the center of the ring. 

As evidenced in Fig.~\ref{Fig5}, a tertiary bubble cloud such as the one observed in Fig.~\ref{Fig2}, appears on a time scale following the reflection of the diverging inner shock at the annular laser-induced bubble wall. These observations suggest that the rebounding shock wave becomes negative upon reflection at the annular laser-induced bubble wall due to the acoustic impedance mismatch between the liquid and the bubble, which leads to the nucleation of the tertiary bubble cloud. We also observe that the nucleated bubble clouds are {\color{black}larger} for smaller laser excitation rings, see the frames bounded by dashed lines in Fig.~\ref{Fig5}. For instance, for the laser ring radius $R$ of 55~$\mu$m and 68~$\mu$m, the bubble clouds fully fill the space enclosed by the annular laser-induced bubble. {\color{black}Qualitatively, it is obvious from the selected images in Fig.~5 that the bubble cloud fills more and more the whole ring as the ring size decreases.} We also speculate that the annular laser-induced bubble increases the lifetime of the nucleated bubble clouds \cite{lohse} by shielding the microbubbles from the liquid static pressure. It seems as well that multiple bubble clouds are nucleated as the reflected inner tensile shock focuses towards the center. Most probably, the bubble clouds in Fig.~\ref{Fig5} have a sufficiently long lifetime for several bubble clouds to appear simultaneously while the shock bounces back and forth inside the annular laser-induced bubble.

\section{Numerical Simulations}

Our numerical simulations intend to describe the inner shock wave propagation and focusing within the laser-induced bubble ring to model the conditions of the appearance of single-bubble and bubble clouds arising from the tensile component of the inner shock wave. Additional effects are not considered in our simulations, including shielding by the annular laser-induced bubble \cite{lohse} and possible shear-induced nucleation due to the interaction between the shock and the glass boundary \cite{Seddon2012, Ando2012, Leora}. {\color{black}The effect of liquid impurities and liquid confinement at different layer thicknesses, that can most probably influence the pressure thresholds for bubble nucleation through heterogeneous nucleation are not accounted in our model.}

We simulate the shock wave evolution by solving a single component Euler equation with the stiffened equation of state for water \cite{stiffened}, assuming for simplicity infinite acoustic impedance mismatch between water and the glass substrate. {\color{black}Since the shock speed is several times larger than the speed of the bubble wall, the bubble nucleation can be considered as quasi-static at the shock timescale and a single component solver is sufficient. The crucial point of the numerical simulations is indeed to model accurately the shock pressure profile during propagation and focusing.} We run a series of numerical simulations by using the \emph{Mezcal} code, an Eulerian code which integrates the hydrodynamics equations by a second order, in space and time, Godunov method \cite{decolle2005, decolle2008}. The code has been extensively used to study fluid dynamics problems. The hydrodynamics equations are integrated in the conservative form by solving the equations regulating the evolution of mass, momentum, and total energy $e$ defined as the sum of thermal $e_{\rm th}$ and kinetic energy $e_{\rm k}$, i.e. $e = e_{\rm k}+e_{\rm th}$. The thermal energy is related to the fluid pressure $p$ by the following equation of state,
\begin{equation}
 e_{th} =
      \frac{p}{\Gamma-1} + \Gamma p_\infty \;,
\end{equation}
with $p_\infty= 3.07 \times 10^8$ Pa, and the adiabatic index $\Gamma = 7.15$ \cite{quinto2013, shyue1998}. The speed of sound is then defined in the code by $c_s=\sqrt{\Gamma (p+p_\infty)/\rho}$.

  \begin{figure}[t!]
 \includegraphics[width=8.5cm]{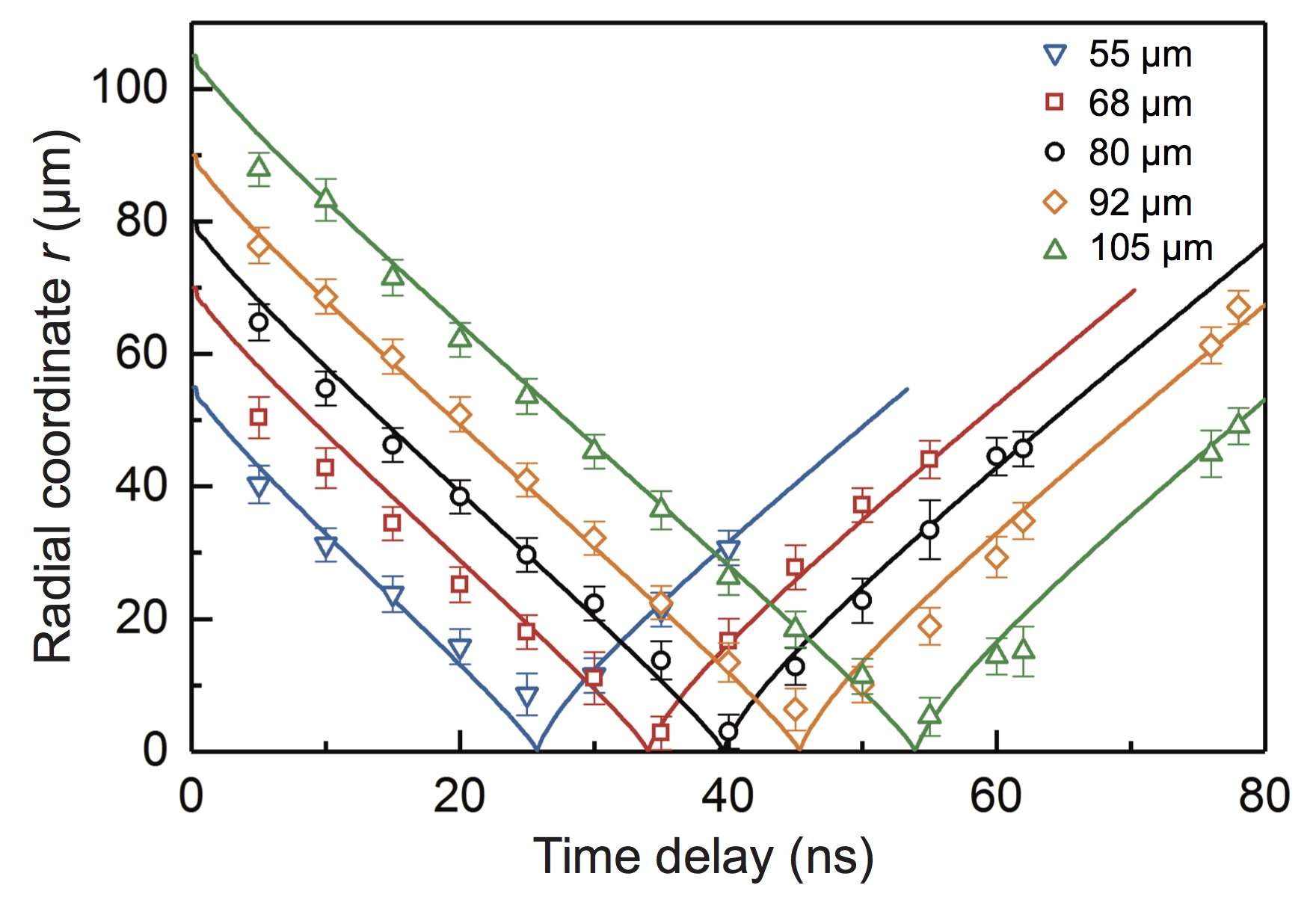}
   \caption[Fig6]{Numerical simulations of the inner shock trajectories for several laser ring radii of $R_{sim} = 55, 70, 80, 90, 105 ~\mu$m and an initial pressure of 2 GPa. The symbols correspond to the inner shock trajectories extracted from Fig.~4 for different laser ring radii.}
   \label{Fig6}
   \end{figure} 

   \begin{figure}[t!]
 \includegraphics[width=8.5cm]{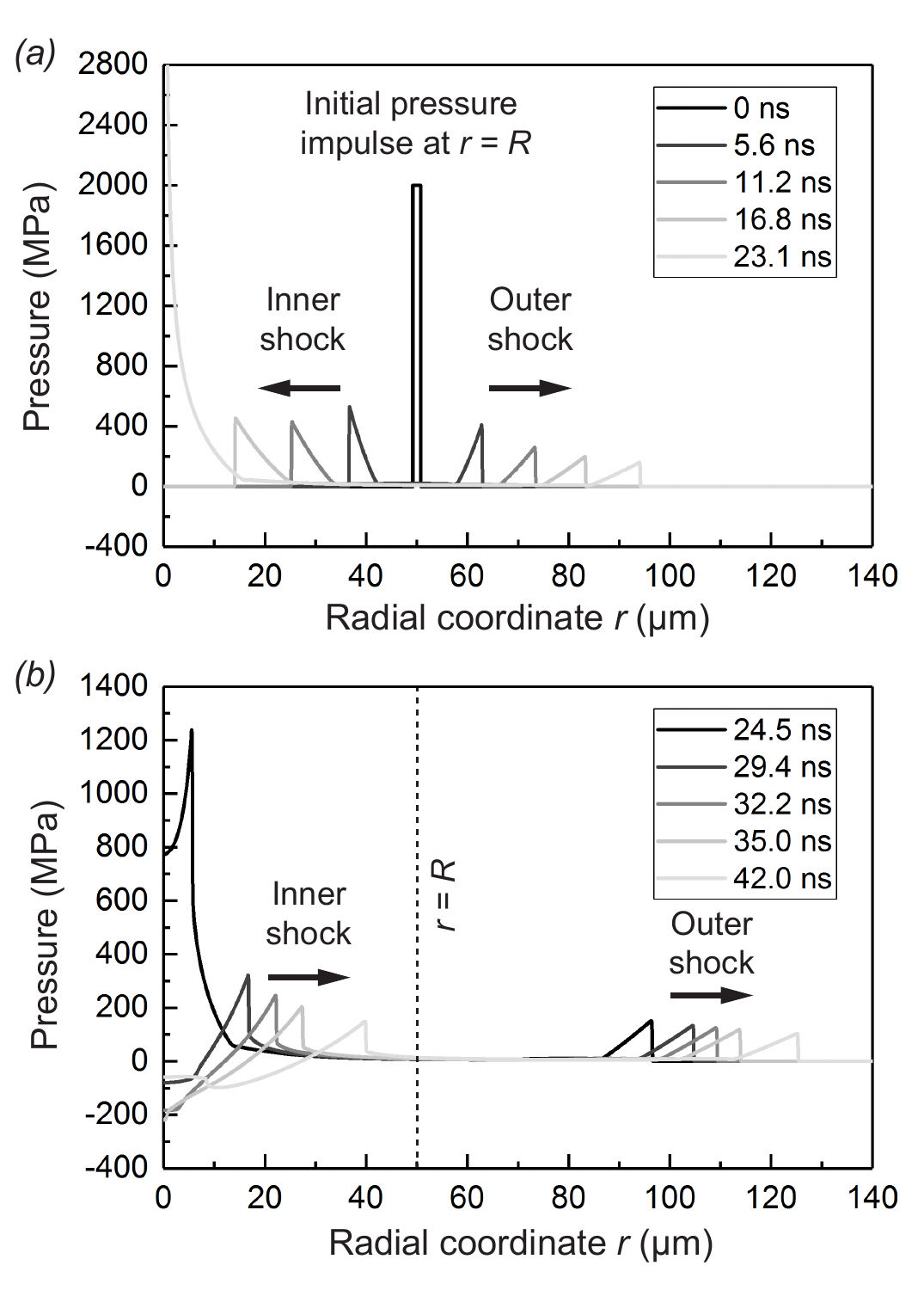}
   \caption[Fig7]{Numerical simulations of the spatial pressure profiles at different times during shock propagation of both, inner and outer shock waves, departing from the laser ring coordinate $r$ = $R$ = 50~$\mu$m. (a) The top figure shows both spatiotemporal pressure profiles until the time of shock focusing. (b) The bottom figure shows the spatiotemporal pressure profiles of both inner and  outer shock waves at later times after shock focusing. The appearance of a tensile tail on the inner shock spatial profile, right after shock focusing, arise from the acoustic discontinuity at the center.}
      \label{Fig7}
   \end{figure} 

The simulations include a one-dimensional, cylindrically-symmetric uniform grid with radial coordinate $r$ in the range of 0-150 $\mu$m with steps of $5~\mu$m resolved by 6000 cells, corresponding to a spatial resolution of $2.5\times 10^{-2}$ $\mu$m per cell. The computational region is initialized by setting an uniform density of $1000$ kg/m$^3$ and a pressure of $10^5$ Pa. We assume that the energy of the laser beam is homogeneously deposited on a ring of radius $R$ and width $\Delta R$ = 1.6~$\mu$m, similar to the diffraction limited spots of the microscope objectives used in this study. We initialize the impulse pressure to a value $p$ = 2~GPa to interpret our experimental observations of Fig.~\ref{Fig5}. We run several models with $R$ varying from 15 to 110~$\mu$m. The evolution of the shock is followed during 1400~ns. 

The simulations of the trajectories of the inner shock waves for different laser ring radii are plotted in Fig.~\ref{Fig6}. The continuous lines represent the results of simulations for several laser excitation radii $R_{sim}$ of 55, 70, 80, 90, and 105~$\mu$m and an initial pressure of 2 GPa while the symbols correspond to the radial position $r$ of the inner shock waves, as they propagate toward and later away from the center ($r$ = 0 $\mu$m), extracted from the images displayed in Fig.~\ref{Fig4}. {\color{black} The error bars on each symbol indicate the uncertainties in tracking the shock front coordinates.} There is {\color{black} a reasonable agreement} between the numerical simulations and the experimental results which confirms the accuracy of the modeling.  {\color{black}We have run additional numerical simulations of a weak shock wave at much longer distances to compare our numerical results to the analytical solutions for the far field acoustic radiation of a ring piston in cylindrical coordinates \cite{Blackstock}. The numerical simulations agree really well with the analytical theory, see supplemental, and suggest that our numerical modeling is  accurate for the modeling of the shock propagation from low to strong amplitudes shock waves.}

The simulations of the time evolution of the spatial pressure profiles for a laser ring radius $R$ of 50~$\mu$m, from $t$ = 0 ns, corresponding to the initial pressure impulse of 2~GPa driven by the laser to the instant at which the inner shock focuses at the center are shown in Fig.~\ref{Fig7}(a). The joint effects of the spatial overlap of the inner shock wave at the center of the ring and the geometrical in-plane confinement of the cylindrical shock wave while approaching the center, entail a sudden and giant increase of the shock pressure right at the shock focus at $r$ = 0~$\mu$m to 40 GPa (the vertical range is limited to 2.8 GPa in Fig. \ref{Fig7}(a)). On the other hand, the geometrical divergence of the outer shock wave induces a gradual decrease in amplitude. Figure~\ref{Fig7}(b) shows the simulated shock pressure profiles at longer times after the inner shock has focused and diverged away from the center. The acoustic discontinuity at the shock focus, an occurrence of the Gouy phase shift \cite{pezeril2011}, is responsible for the transformation of the unipolar shape of the incoming spatial shock profile into a bipolar spatial shock profile with a tensile pressure tail. The simulations in Fig.~\ref{Fig7}(b) confirm that upon crossing the shock focus, the inner shock profile becomes bipolar with a characteristic positive pressure front and a tensile pressure tail. The tensile pressure tail, which in the equation of state corresponds to a negative pressure, can stretch the liquid below vaporization resulting in bubble cavitation right at the shock focus where the tensile pressure is maximum, as evidenced in our experimental observations of the appearance of a central bubble at the shock focus. In the simulations displayed in Fig.~\ref{Fig7}(b), we also notice that after reaching a maximum tensile value at the center at $r$ = 0~$\mu$m, the tension at the center decreases as the shock wave propagates away, and the minimum value of the pressure shifts and extends to larger values of $r$, see Fig.~\ref{Fig7}(b) at 42 ns, which indicates that the bubble cavitation effect at the center can probably spread at long distances to the focus.

\begin{figure}[t!]
 \includegraphics[width=8.5cm]{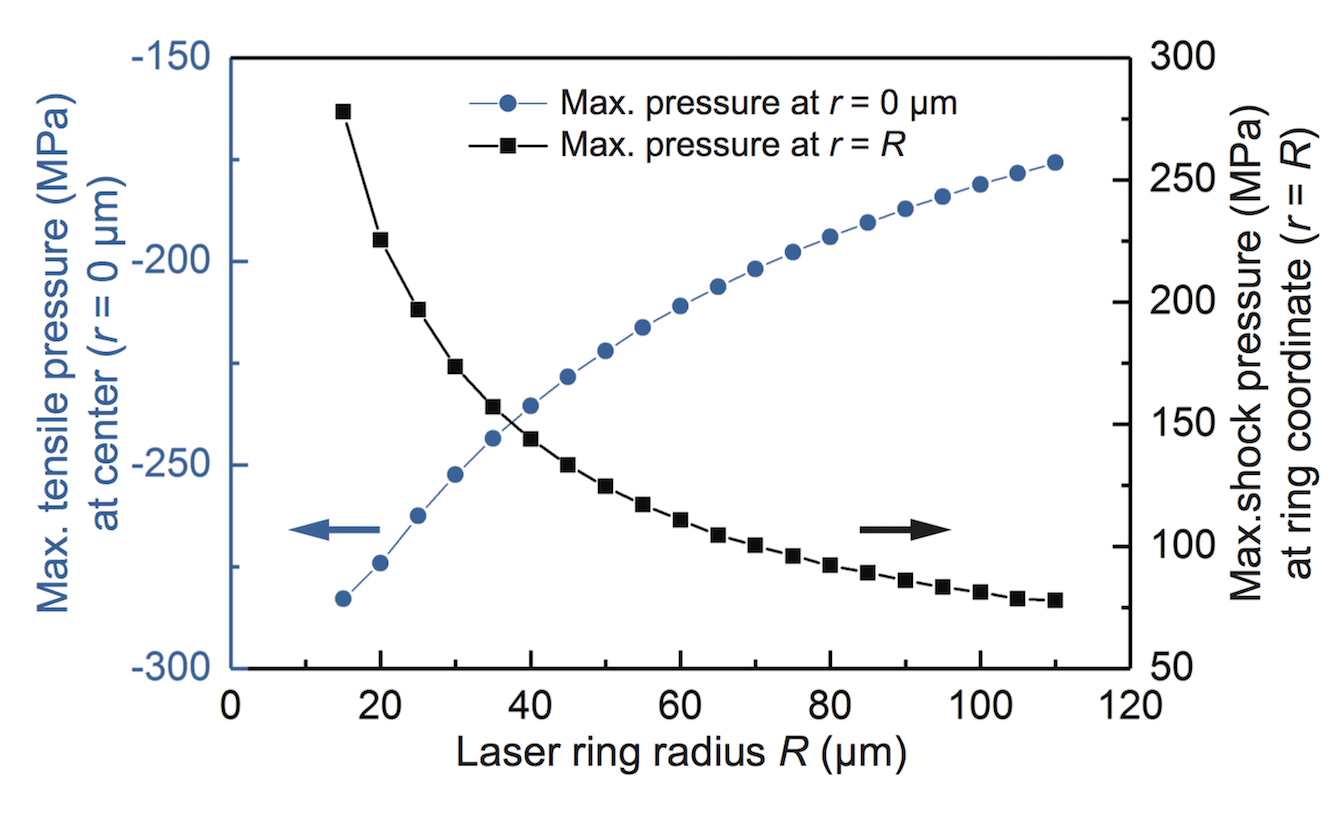}
   \caption[Fig8]{(left axis) Maximum tension pressure as a function of the laser ring radius $R$, these are reached at the center ($r$ = 0~$\mu$m) shortly after the inner shock focuses. (right axis) Maximum positive pressure as the inner shock rebounds and reaches original position $r$ = $R$. }
   \label{Fig8}
   \end{figure}

In order to interpret the behavior of the appearance of the central bubble, we have performed numerical simulations of the tensile pressure at the center for different values of the laser excitation radius $R$. The plot of the largest tensile pressures reached at the center at $r$ = 0~$\mu$m as a function of the laser excitation radius $R$ is shown in Fig.~\ref{Fig8}(left axis). The maximum tensile pressure is reached for the smallest $R$ and decreases monotonically for larger values of $R$. In the experiment, the liquid is expected to break at moderate negative pressures due to heterogeneous nucleation. Therefore, the central cavitation bubble should appear as soon as the value of the pressure drops below the vapor pressure, which occurs a few nanoseconds after shock focusing. As seen in the simulations of Fig.~\ref{Fig8}(left axis), the tensile pressure tail of the focusing shock wave is well below the vapor pressure and should induce bubble cavitation at the center for any given $R$, however, the tensile pressure being higher for small $R$, the bubble cavitation effect at the center is expected to be more efficient for small $R$. {\color{black}It is to note that our simulations predict tensile pressures at the center of focus in the range of hundreds of MPa, which is well above heterogeneous or homogenous cavitation thresholds. Because the present method is based on focusing and pressure amplification, it is particularly appropriate for creating large tensile transient pressures but can hardly serve as a way to measure nucleation thresholds. Indeed, weak shocks, obtained by using lower laser energies or larger excitation radii, approach the linear acoustic limit and pressure estimations can no longer be deduced from the experimental measurement of the acoustic speed, contrary to \cite{pezeril2011}.}

Similarly, in order to model the conditions of the appearance of the bubble cloud, we have performed numerical simulations of the inner shock wave propagation away from the shock focus, until it reaches the annular laser-induced bubble wall where it gets converted into a tensile wave. Figure~\ref{Fig8}(right axis) shows the maximum value of the pressure of the inner shock front right at the laser ring coordinate $r=R$, where it gets reflected by the laser-induced annular bubble and converted into a tensile wave, for different values of the laser excitation radius $R$. For simplicity, we assume that the laser-induced annular bubble does not expand. From the simulations displayed in Fig.~\ref{Fig8}(right axis), it appears that smaller radius $R$ lead to higher shock pressures at the ring coordinate. Hence, we expect stronger reflected tensile shock waves for smaller laser radii which should most likely give rise to larger bubble clouds. This is qualitatively confirmed by our experimental observations of bubble clouds for different radius $R$ in Fig.~\ref{Fig5}, where larger tertiary bubble clouds are observed for smaller laser excitation rings.\\

\section{Conclusion}

We have experimentally observed several transient phenomena such as single-bubble cavitation as well as bubble cloud nucleation as a result of the propagation and focusing of a cylindrical shock wave.

Our experimental results are supported by numerical modeling, which have shown that, as expected, the rebounding inner shock stretches the liquid at the center of the ring. This results in the appearance of a single cavitation bubble at the center which expands and collapses in a few microseconds. The nucleation of a tertiary bubble cloud, resulting from the reflection of the inner diverging shock at the annular laser-induced bubble, is experimentally observed and is supported by our numerical modeling as well. The experimental observations of larger bubble clouds for smaller laser ring radii agree qualitatively with our simulations. Our findings shed light on shock-induced cavitation and bubble nucleation.

Ultrasound-driven sonoluminescence in confined liquid geometries like microfluidic channels has been observed in \cite{ohl}. So far our experimental attempts for the observation of sonoluminescence in our specific cylindrical configuration and liquid confinement of bubble implosion has been unsuccessful. At the moment, the present work of converging shock-driven bubble creation and implosion opens novel perspectives and challenges in the frame of high amplitude shock waves in liquids and energetic cavitation \cite{Bell, Vogel, Sankin, antonov, niteshinskiy, Ramsey}.

\section*{Acknowledgments} 
Work partially supported by the following grants:
U.S. Army Research Office through the Institute for Soldier Nanotechnologies, contract number W911NF-13-D-0001,
Office of Naval Research DURIP grant number N00014-13-1-0676, CONACYT contract number 253706, 253366, 269314, DGAPA-UNAM, (PAPIIT) contract number IN104415, IN117917, IA103315, and CNRS (Centre National de la Recherche Scientifique) under PICS contract (Projet International de Coop\'eration Scientifique).

\end{document}